# From Open Source Intelligence to Decision Making: a Hybrid Approach

Vitaliy Tsyganok[1][0000-0002-0821-4877], Sergii Kadenko[1][0000-0001-7191-5636],

Oleh Andriichuk[1] [0000-0003-2569-2026]

[1] Institute for Information Recording of the National Academy of Sciences of Ukraine, 03113, 2, Shpak str., Kyiv, Ukraine

**Abstract.** We provide an overview of tools enabling users to utilize data from open sources for decision-making support in weakly-structured subject domains. Presently, it is impossible to replace expert data with data from open sources in the process of decision-making. Although organization of expert sessions requires much time and costs a lot, due to insufficient level of natural language processing technology development, we still have to engage experts and knowledge engineers in decision-making process. Information, obtained from experts and open sources, is processed, aggregated, and used as basis of recommendations, provided to decision-maker. As an example of a weakly-structured domain, we consider information conflicts and operations. For this domain we propose a hybrid decision support methodology, using data provided by both experts and open sources. The methodology is based on hierarchic decomposition of the main goal of an information operation. Using the data obtained from experts and open sources, we build the knowledge base of subject domain in the form of a weighted graph. It represents a hierarchy of factors influencing the main goal. Beside intensity, the impact of each factor is characterized by delay and duration. With these parameters taken into account, main goal achievement degree is calculated, and changes of target parameters of information operation object are monitored. In order to illustrate the suggested hybrid approach, we consider a real-life example, where we detect, monitor, and analyze actions intended to discredit the National academy of sciences of Ukraine. For this purpose, we use specialized decision-making support and content monitoring software.

## 1 Introduction: problem outline and current state of research in the area

In recent years the problem of aggregating data coming from different types of sources has been rapidly gaining relevance. On the one hand, we are talking about data, obtained from experts (competent specialists in narrow subject domains), on the other – data from printed and electronic publications, data bases of documents, reports, as well as blogs and even comments in social media. In order to hold an expert session (even in remote mode), one needs to find and involve at least several experts, pay for their work, and dedicate some organizational efforts to the process. At the same time, analysis of "non-expert" data from open sources (OS) is not associated with any direct expenses, because these data are publicly available, and the knowledge engineer (assessor, moderator) or decision-maker (DM) can always use them. During the last few decades many powerful approaches and methods were developed for analysis and processing of expert data of various formats. These approaches became the basis of mathematical ware of numerous decision support systems DSS. In view of these considerations, it would be appropriate and promising to try to extrapolate existing expert data analysis and processing means and methods to data, coming from OS [1]. This will allow us to save resources and efforts on expert session organization and make decision-making support process more affordable and flexible. At present, it is too early to speak of completely refraining from usage of expert data in decision-making process and its replacement with data from OS. However, the task of developing a methodology that would be based on simultaneous usage of both expert data and data from OS seems quite manageable.

As we know, expert methods are used for decision-making support in the so-called weakly structured subject domains, where determined data is scarce [2]. Examples of such subject domains include forecasting, competitive project selection, sustainable development, environmental protection [3], and space industry [4]. Weakly structured character of information warfare sphere (including information operation (IO) detection and counteraction), is confirmed by a set of factors ([5], [6]). Information impact upon conscience of the target audience is exerted, to a large extent, through OS (websites, blogs, social media, electronic and printed publications, TV etc). So, based on OS data (if properly analyzed and processed) we can compile formal description of a specific IO, evaluate its outcomes, and make a decision as to counteracting it. Modern sources feature lots of publications, dedicated, on the one hand, to information warfare ([7] - [9]), textual analytics ([10], [11]), and detection of false information intended for manipulating public opinion [12], and on the other – to methods of decision-making support during strategic planning [13] and analysis of public preferences [14]. However, no significant attempts to combine the means of expert data and OS data processing in the information warfare context have been made so far.

The purpose of the current research is to analyze the prospects of OS usage for decision-making in general, and for information warfare in particular, as well as to develop a hybrid methodology for decision-making support during recognition of IO, based on both expert and OS data.

## 2   Definition and role of open information sources

According to Army Techniques Publication [15], an OS is a source of information that provides it without the need to preserve its confidentiality, i.e., provides infor-mation that is not protected from public disclosure. According to a research, conducted by Delphi Group in early 2000-s [16], the relative weights of different information sources are as shown on Fig. 1. However, in recent years these weights have changed as a result of increasing volumes of information becoming available from printed and, especially, digital OS.

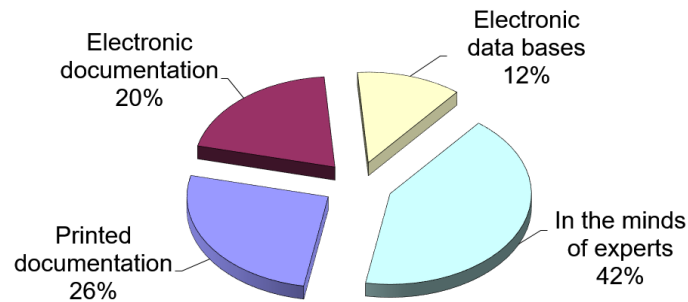

**Fig.1** Relative weights of information sources

The key directions of OS usage for decision-making support include analytical research, priority-setting in activity of organizations and DM, distribution of limited resources among projects (competitive selection of projects for funding), and development of strategic plans. Strategic planning technology, in its turn, includes the following phases [4]:

- Formulation of the main strategic goal.
- Building of a hierarchy of criteria (a system of indicators) that characterize the subject domain (*potentially – using data from OS*).
- Definition of the relative weights of criteria (projects) in the hierarchy (*potentially – using data from OS*).
- Calculation of relative efficiency of projects, i.e. "relative contribution" of each project into achievement of the main goal.
- Definition of an optimal strategy as the most efficient distribution of limited resources among projects (from the standpoint of the formulated main goal).

The common purpose of all the listed directions is *to provide competent and informative recommendations to the DM concerning his further activities*.

Peculiar features of working with OS in the process of decision-making support are as follows.

1) Advantage: opportunity to save resources needed for expert session organization. Information from OS is publicly available, so it allows us to avoid loss of efforts and resources on expert session organization.

2) Advantage: larger volume of available information. Usually, only a few experts participate in the session. The information they provide is often incomplete. OS provide large volumes of data, so the knowledge engineer or the DM has better chances of getting the information he needs.

3) Disadvantage: way of data representation. We can expect (or request) an expert to provide data in the format that simplifies its input into the knowledge base (KB) of the DSS. In the ideal case expert estimates are provided in the specified scale (ordinal or cardinal). If the expert feels uncomfortable inputting data in numeric format, we can offer him to choose the respective verbal equivalents of pair-wise comparison scale grade values, or hold a coaching session before the expert examination [17]. However, in the case of a generalized OS, knowledge engineer has to deal with the format, in which the data are provided in the source, and manually bring the data to the format acceptable by the DSS KB (unless this function is automated).

4) Disadvantage: impossibility of feedback. If data obtained from experts is not sufficiently complete, consistent, compatible, or detailed, there is an opportunity (at least, in theory) to improve the quality of the data through feedback with experts (i.e. requesting them to reconsider their judgments). In the case of an OS, such feedback is impossible, so it makes sense to resort to approaches to definition of relative weights of data sources based on the listed qualities [18].

Examples of data from OS that can be used for decision-making support include (but are not limited to) the following ones. Publications of complete ratings and rankings (that are themselves the result of pair-wise comparisons, polls, surveys, or direct estimation sessions) can be used in DSS as ready-made direct estimates. Results of work of several search engines can be viewed as individual alternative rankings. Aggregate ranking can be used as basis for recommendations concerning decision variant choice [19]. If there is a need to compare two objects according to a certain criterion, we can compare the frequencies of references to these objects in OS. In this process, it is necessary to consider the context of reference (positive, negative, or neutral) ("analyze the sentiment") and credibility of sources. Even the very number of references to some object can serve as a sort of rating (in the case when the frequency of reference is really proportionate to the object's significance according to the criterion). An example, where frequency of references to objects is used as a source of data for DSS KB, is provided below. Another example – references to pairs of objects in comparative context. In case of ordinal pair-wise comparisons we should look for references, including such verbal expressions as "A is *better/worse* than B". In case of cardinal comparisons, in addition to ordinal comparison, we should look at preference degree (for instance, "*A is slightly/moderately/significantly better/worse than B*": very similar verbal equivalents are used by Saaty and Likert [17] in their respective scales).

Available resources, means, and tools that can be used for decision-making support based on OS data, include the following ones: the OS themselves and web search engines; textual analytics tools (for instance, those embedded in InfoStream content monitoring system [1], [20], Textual Analysis Suit module of SAS Text Mining system [21]); DSS ("Consensus-2" [4], "Solon" [3], [4], "Riven' (Level)" [3], SuperDecisions [22], "Svir'" [23]). Based on available arsenal of resources and tools, conceptual model of OS data usage for decision-making support, will look as shown on Fig. 2.

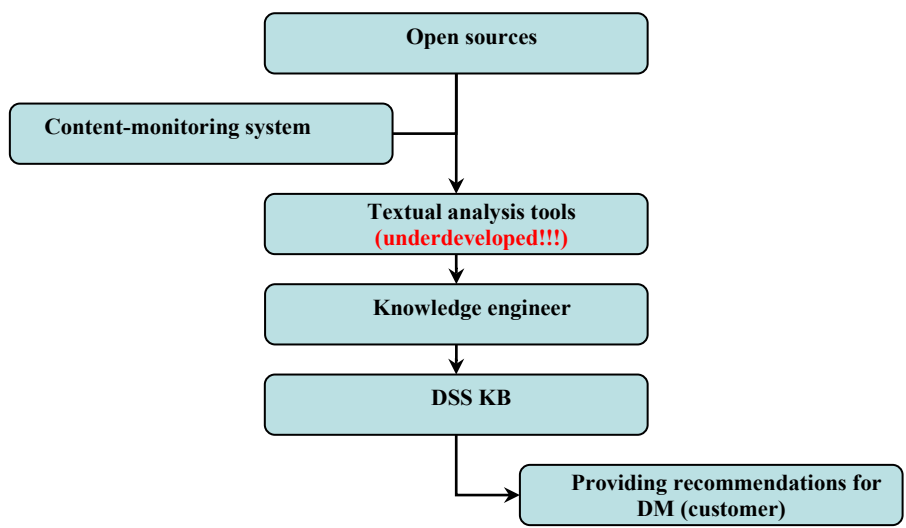

**Fig. 2** Conceptual model of the process of OS data usage for decision-making support

Due to insufficient level of natural language processing and sentiment analysis tools development, the respective operations are performed, mostly, manually, by moderators, assessors, and knowledge engineers. As we see, available arsenal of modern DSS and content-monitoring systems (CMS) does not allow us to thoroughly automate such processes as data search for decision-making and input of this data into DSS KB for further processing and formulation of recommendations for DM. So, at present we can use combined (hybrid) approaches to decision-making, using information coming from both experts and OS. In the hybrid methodology, described below, expert information is used for DSS KB construction, while OS information is used to "fine-tune" the DSS KB structure and append its content.

## 3  A Hybrid Decision Support Methodology for IO Recognition

Application of the hybrid methodology we propose to IO recognition envisions the following sequence of steps.

1. Conduct preliminary research of IO object and select its target parameters (indicators). We assume that IO against the object took place before (in retrospect), and resulted in worsening of its state/condition (i.e. respective target indicators).

2. Hold group expert sessions in order to define and decompose IO goals, as well as estimate the degree of impact of different IO components upon the IO object. IO is decomposed as a complex weakly structured system. For this purpose we are using the tools of the system for distributed collection and processing of expert information (SDCPEI). To obtain thorough and undistorted expert knowledge, it makes sense to use the respective expert estimation system (EES).

3. Build the respective KB of the subject domain. This step is performed using DSS tools based on the results of conducted group expert session and available objective data.

4. Analyze the dynamics of topical information stream using the CMS tools. DSS KB is appended.

5. Calculate recommendations for the DM (using DSS tools and constructed KB). Calculate IO goal achievement degrees in retrospect and analyze their correspondence with the respective changes of IO object state/condition. Calculate average values of IO goal achievement degrees, under which the target indicators of IO object became worse.

Thus, through monitoring of the IO object states during the current time period, we can forecast worsening of IO object's target indicators, based on comparison of IO goal achievement degrees, calculated for the current period, with the respective average values. If we have statistically sufficient set and sufficient correlation between IO goal achievement degrees and worsening of target indicators of the IO object, we can even forecast numeric value of worsening of target indicators of IO object for the present moment.

The advantages of the suggested methodology are as follows.

1. Considerable degree of detail of the model: against the background of a large number of publications about the IO object in general, amplitudes of oscillations, induced by disinformation targeted at its particular components might seem insignificant, and, consequently, hard to detect. The methodology allows us to overcome this drawback through hierarchical decomposition of the main goal of the IO.

2. Larger data set can be monitored, as there are more queries and more keywords.

3. Weighting of IO components (assigning of relative importance coefficients) allows us to avoid the situation when all components have the same weight. As a result, an adequate IO model can be built.

4. Once built, the KB can be used for a considerable period of time in future, without the necessity to hold expert sessions again.

5. Thanks to SDCPEI tools, experts can work online, so we can save time and resources on expert sessions.

The disadvantages of the suggested methodology are as follows.

1. Usage of expert technologies still requires considerable time and funds, needed to pay for group expert sessions, as well as timely updates of the KB for its future uses.

2 Complex and, sometimes, ambiguous character of formulation of some difficult-to-understand IO components as queries submitted to CMS.

## 4 Example

Let us illustrate the methodology, outlined in the previous section, by an example, where we analyze the process of discrediting of the National academy of sciences (NAS) of Ukraine in the information space. During the last years funding of the NAS was reduced several times. Percentage of expenditures allocated for the activity of the NAS of Ukraine (NASU) in the state budget also decreased. This can be confirmed by the data on the distribution of expenses of the State budget of Ukraine for 2014-2016. It would be reasonable to assume that, among other reasons, such reduction of funding volumes results from discrediting of the NASU in the media. Discrediting of the NAS is beneficial for people and organizations that compete with the academy for budget funding and promote certain decisions regarding the academy's institutions and assets. Expert examination targeted at detection of potential IO would allow us to forecast further development of the situation and provide interested parties with recommendations on counteracting the IO and influencing the target indicators of the IO object (i.e. the NAS).

On steps one to three of the hybrid methodology the hypothetical main goal "IO against the NAS of Ukraine" has been decomposed (by experts) and KB of the respective subject domain has been built. For these tasks the tools of SDCPEI "Consensus-2", EES "Riven'" ("Level"), and DSS "Solon-3" ([3], [4]) were used. The experts included employees of different NAS institutions. The KB obtained as a result of the respective steps of the methodology, was represented in the form of a weighted hierarchy graph, including sub-goals of the main goal. As a result of decomposition of the main goal by the expert group, 15 formulations of main goal components were obtained. The components are as follows: 1) bureaucracy in the NAS of Ukraine; 2) inefficient staff policy of the NASU; 3) corruption in the NASU; 4) understatement of the level of scientific achievements of the NASU; 5) lack of implementation of academic research results into production practice; 6) understatement of international cooperation level; 7) improper and inefficient usage of NASU property; 8) improper and inefficient usage of NASU land resources; 9) discrediting of the president of the NASU; 10) discrediting of the executive manager of the NASU; 11) discrediting of other renown NASU representatives; 12) contraposition of scientific achievements of the NAS and the Ministry of education and science; 13) contraposition of scientific achievements of the NAS and other academic institutions; 14) contraposition of achievements of Ukrainian companies and the NASU; 15) contraposition of scientific achievements of foreign institutions and the NASU.

Next we move to the fourth step of the methodology, where we perform the analysis of dynamics of topical information flow using the tools of InfoStream CMS [1], [20]. For this purpose, queries are formulated on a specialized language (in accordance to every listed component of potential IO). According to these queries, analysis of dynamics of publications on target topics should be performed.

Fig. 3 illustrates the results of express-analysis of topical information flow [24] which corresponds to the potential IO object, i.e. NASU. As a result of express-analysis, using the tools of InfoStream CMS, we obtained the respective topical information flow from Ukrainian web segment. To detect information "throw-ins", using the available tools, we analyzed the dynamics of publications on target topics. The figure illustrates publication dynamics for the period between 01.07.2015 and 31.12.2015.

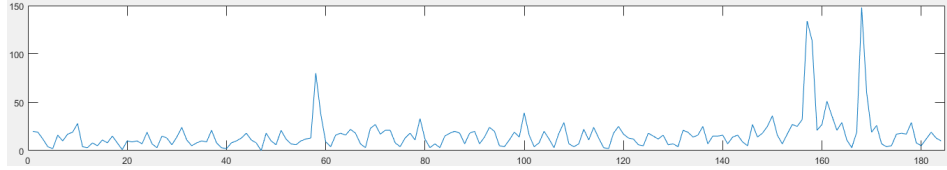

**Fig. 3** Dynamics of publications on the target topic "Inefficient staffing policy of the NASU"

In order to define the degree of similarity between fragments of the respective time series and the diagram of a typical IO in different scales, we are using wavelet analysis. A wavelet is a function that is well-localized in time and can be used as behavior template on a small-to-medium time interval.

Wavelet coefficients indicate the degree of similarity between the process's behavior in a certain point and a typical wavelet of a given scale [25]. On the respective wavelet spectrograms (Fig. 4) we can see all characteristic features of the initial time series: scale and intensity of periodic changes, direction and values of trends, presence, location, and duration of local phenomena. X axis represents time (in days) since the beginning of the interval under consideration. Y axis represents wavelet scales. The color indicates the intensity of the wavelet (light yellow means the highest intensity, while dark blue – the lowest).

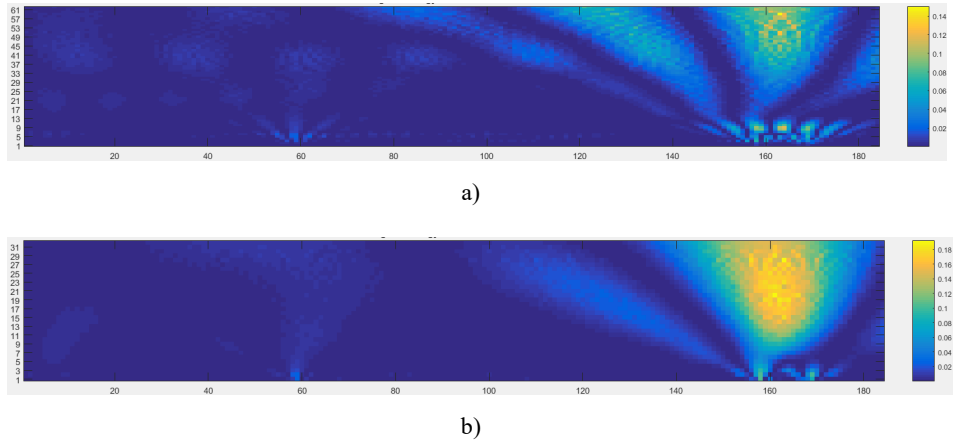

a)

b)

**Fig. 4** Wavelet spectrograms of the information flow: a) Morelet wavelet; b) "Mexican hat"

Our choice of Morelet (Fig. 4a) and "Mexican hat" (Fig. 4b) wavelets can be explained by the fact that they most precisely represent the dynamics of IO [25], [26], [1]. That is why we analyze time series, corresponding to each of the 15 potential IO components during 7 time intervals (01.01.2013-31.12.2013, 01.01.2014-31.12.2014, 01.01.2015-31.12.2015, 01.01.2016-31.12.2016, 01.01.2017-31.12.2017, 01.01.2018-31.12.2018, та 01.01.2019-10.07.2019) and try to identify the presence of the listed two wavelets.

At the fifth step of the methodology, based on information impacts or "throw-ins", detected on the previous step, and their parameters (duration and location on the time axis), the knowledge engineer supplements (appends) the KB of the DSS "Solon-3". Particularly, in the example under consideration, we have identified an informative "throw-in" concerning the potential IO component "Understatement of the level of scientific achievements of the NASU", located on 30.11.2015 on the time axis and lasting 14 days. Respectively: 1) as a characteristic of the project "Understatement of the level of scientific achievements of the NASU" we introduce a parameter of project duration, that equals 14 days; 2) as a characteristic of the project's impact upon the goal "Discrediting of the scientific achievements of the NASU" we introduce a parameter of impact delay for the term of 10 months. For all other detected informative "throw-ins" we introduce similar characteristics in a similar way. So, for the period 01.01.2015-31.12.2015 the appended KB assumes the structure, displayed on Fig. 5. Table 1 contains the list of formulations of all goals and projects of the KB.

We should note that for each of the following potential IO components – "Corruption in the NASU", "Bureaucracy in the NASU", "Inefficient staff policy of the NASU", "Improper and inefficient usage of NASU land resources", and "Improper and inefficient usage of NASU property" – 2 information "throw-ins" were detected during 2015. That is why the respective projects were input into the data base 2 times each. For instance, for potential IO component "Bureaucracy in the NASU", two projects were input: "Bureaucracy in the NASU 1" and "Bureaucracy in the NASU 2". They have different durations (9 and 15 days), while their impacts have different delays (9 and 11 months), respectively.

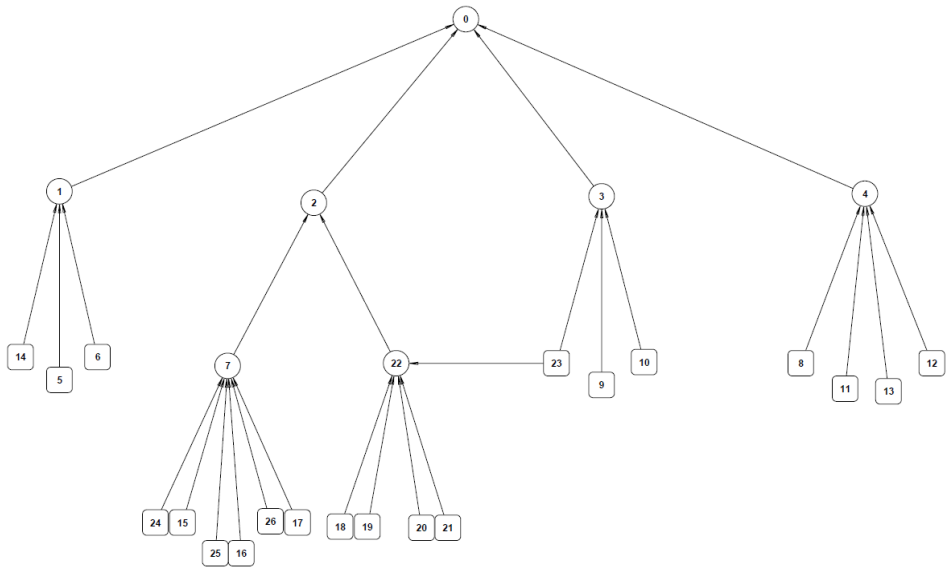

**Fig. 5** Structure of the KB built in "Solon-3" DSS

**Table 1** List of goal formulations

| # | Goal formulation |
|---|---|
| 0 | IO against the National academy of sciences of Ukraine (NASU) |
| 1 | Discrediting of the scientific results obtained by the NASU |
| 2 | Discrediting of the institutions of the NASU |
| 3 | Discrediting of renown representatives of the NASU |
| 4 | Overstatement of academic achievements of the institutions competing with the NASU |
| 5 | Lack of implementation of academic research results into production practice |
| 6 | Understatement of international cooperation level |
| 7 | Discrediting of the organizational structure of the NASU |
| 8 | Contraposition of scientific achievements of the NAS and the Ministry of education and science |
| 9 | Discrediting of the president of the NASU |
| 10 | Discrediting of other renown NASU representatives |
| 11 | Contraposition of scientific achievements of the NAS and other academic institutions |
| 12 | Contraposition of scientific achievements of foreign institutions and the NASU |
| 13 | Contraposition of achievements of Ukrainian companies and the NASU |
| 14 | Understatement of the level of scientific achievements of the NASU |
| 15 | Corruption in the NASU 2 |
| 16 | Bureaucracy in the NASU 2 |
| 17 | Inefficient staff policy of the NASU 2 |
| 18 | Improper and inefficient usage of NASU property 1 |
| 19 | Improper and inefficient usage of NASU property 2 |
| 20 | Improper and inefficient usage of NASU land resources 1 |
| 21 | Improper and inefficient usage of NASU land resources 2 |
| 22 | Discrediting of the executive office of the NASU |
| 23 | Discrediting of the executive manager of the NASU |
| 24 | Corruption in the NASU 1 |
| 25 | Bureaucracy in the NASU 1 |
| 26 | Inefficient staff policy of the NASU 1 |

Next, project implementation degrees are input into "Solon-3" DSS. If for some potential IO components (such as "Contraposition of achievements of Ukrainian companies and the NASU" and "Discrediting of the executive office of the NASU") no information impacts are detected, the respective projects are assigned 0% implementation degrees. For all other projects implementation degrees amount to 100%.

Now we can get the data of recommendation calculations, namely, the implementation degree of the potential IO's main goal (Fig. 6) and project efficiencies.

During the periods 01.01.2013-31.12.2013, 01.01.2014-31.12.2014, 01.01.2015-31.12.2015, 01.01.2016-31.12.2016, 01.01.2017-31.12.2017, 01.01.2018-31.12.2018, and 01.01.2019-10.07.2019 degrees of the main goal achievement amounted to: 0.380492, 0.404188, 0.570779, 0.441362, 0.7597732, 0.5154478, and 0.2443254, respectively. So, in retrospect, the average value of the main goal achievement equals: (0.380492 + 0.404188 + 0.570779 + 0.441362 + 0.7597732 + 0.5154478) / 6.0 = 0.512007.

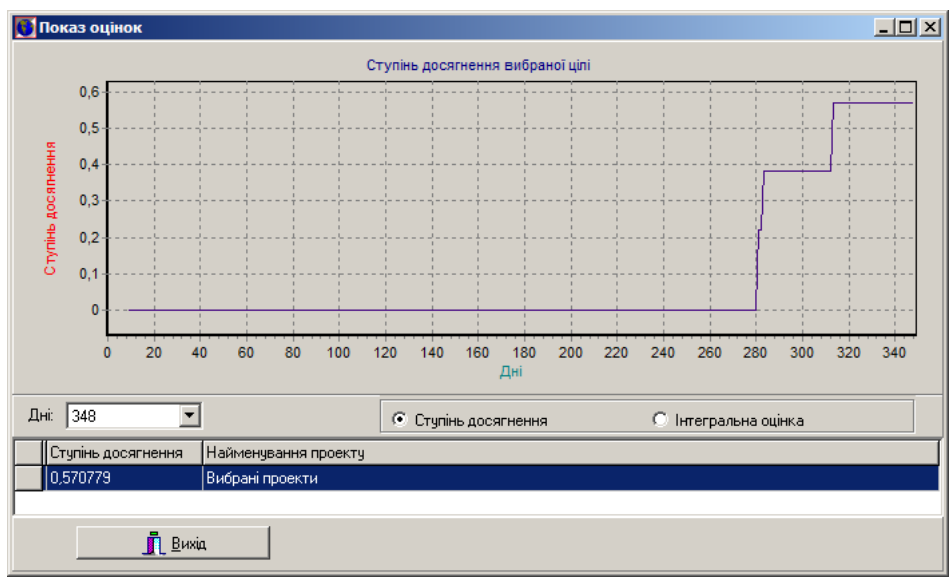

**Fig. 6** Degree of achievement of a selected goal in "Solon-3" DSS

If we assume that during the second half of 2019 the influence of information impacts, directed against the NASU, remains at the same level as it was in the first half of the year, we can expect that at the end of 2019, the degree of potential IO's main goal achievement will amount to 0.2443254×2=0.4886508. This value differs from the value calculated as the average mean of retrospective indicators of the main goal achievement (0.512007) approximately by 4.6%.

Since the average retrospective value of the level of potential IO's main goal achievement and the estimated value expected at the end of 2019 are very close (the difference is less than 4.6%), we can conclude, that during the period under consideration, worsening of the target indicators of the object are, quite probably, caused by an IO.

We should also note, that in 2016-2017 the share of funding allocated to the NASU in Ukrainian budget decreased. Only in 2018 the funding volumes allocated to the NASU were increased by 38%, and this caused another wave of accusations targeted at the academy. Employees of the NASU often played significant roles in these accu-sations, consciously, or unconsciously. Anyway, this funding increase allowed the academy to provide just for the basic level of activity. In June 2019 an open letter of the Council of Presidents of Academies of science of Ukraine to the Prime-minister of Ukraine was published. The letter stressed the critical material condition of the Academies. So, discrediting of the NASU and profile Academies of sciences of Ukraine in the information space goes on, and, consequently, these institutions have to struggle (information struggle included) with competitors and ill-wishers for reputation and sufficient budget funding.

## 5  Conclusions

OS of data are a powerful resource that can and should be used for decision-making support. Active usage of data from OS can help make decision-making process less complex and resource-intensive. To obtain and process the data from OS it makes sense to use CMS jointly with DSS. Available toolkit of modern DSS and CMS does not allow us to completely automate the processes of data search for decision making and of input of these data into DSS KB for processing. Particularly, the functions associated with natural language processing, require improvement and automation. However, even at present development stage, usage of data from OS is a promising direction of decision support technology development, especially, in the information warfare sphere.

In the paper we have demonstrated the relevance of expert data-based decision support toolkit usage in the process of IO identification. We have also suggested a concept of a new information and analytical system for IO recognition, based on systematic integration of SDCPEI, EES, DSS, and CMS.

Besides that, we have proposed and analyzed a hybrid methodology for usage of expert data-based decision support toolkit usage during IO detection, allowing us to forecast the changes of IO object's target parameters in the upcoming period, based on retrospective analysis. The suggested methodology has been illustrated by an example where potential IO targeted against the NASU, is detected and analyzed. Empirical results obtained on real data, confirm the effectiveness of the methodology implementation in the information warfare area.